\documentclass[pra,aps,footinbib,twocolumn,superscriptaddress,citeautoscript,longbibliography]{revtex4-1}

\usepackage[utf8x]{inputenc}
\usepackage{array}
\usepackage{amssymb}
\usepackage{amsmath}
\usepackage{amsfonts}
\usepackage{color}
\usepackage{graphicx}
\usepackage{bm}
\usepackage{dsfont}
\usepackage{url}
\usepackage{braket}
\usepackage{empheq}
\usepackage{tikz}
\usepackage[unicode]{hyperref}
\usepackage{color}
\usepackage{soul}

\hypersetup{
   unicode=true,          
   plainpages=false,
   colorlinks=true,       
   citecolor=blue,        
}
\allowdisplaybreaks
\graphicspath{{.}{./fig/}}

\newcommand{\field}[1]{\mathbb{#1}} 

\newcommand{\rme}{\mathrm{e}}
\newcommand{\rmi}{\mathrm{i}}
\newcommand{\x}{\mathrm{x}}
\newcommand{\y}{\mathrm{y}}
\newcommand{\z}{\mathrm{z}}
\renewcommand{\r}{\mathbf{r}}

\newcommand{\rp}{{\textbf{r}^\prime}}
\renewcommand{\k}{{\textbf{k}}}

\newcommand{\B}{{\mathbf{B}}}
\newcommand{\ua}{\uparrow}
\newcommand{\da}{\downarrow}

\renewcommand{\sp}{{\sigma^\prime}}
\renewcommand{\d}{\ensuremath{\mathrm{d}}}
\newcommand{\zmn}{h} 
\renewcommand{\u}{\mathbf{u}}
\newcommand{\A}{\mathcal{A}}
\renewcommand{\B}{\mathcal{B}}
\renewcommand{\C}{\mathcal{C}}
\newcommand{\D}{\mathcal{D}}

\newcommand{\K}{\mathcal{K}}

\renewcommand{\a}{^{(1)}}

\newcommand{\Pzero}{\mathcal{P}^{(0)}}

\newcommand{\Ac}{\mathcal{A}^{(3)}}

\newcommand{\Af}{\mathcal{A}^{(6)}}

\newcommand{\Ag}{\mathcal{A}^{(7)}}
\newcommand{\Bg}{\mathcal{B}^{(7)}}

\newcommand{\Kg}{\mathcal{K}^{(7)}}

\renewcommand{\c}{^{(3)}}

\newcommand{\f}{^{(6)}}
\newcommand{\g}{^{(7)}}

\newcommand{\nmax}{\nu_{\mathrm{max}}} 

\newcommand{\OP}{\Delta}

\newcommand{\sect}[1]{\textit{#1} -- }
\newcommand{\subsect}[1]{\textit{#1} -- }

\begin{document}

\title{Non-Abelian Majorana fermions in topological $s$-wave Fermi superfluids}
\author{L. A. Toikka}
\email{lauri.toikka@gmail.com}
\affiliation{Institute for Theoretical Physics, University of Innsbruck, A-6020 Innsbruck, Austria}
\date{\today}

\begin{abstract}
By solving the Bogoliubov--De Gennes equations analytically, we derive the fermionic zero-modes satisfying the Majorana property that exist in vortices of a two-dimensional $s$-wave Fermi superfluid with spin-orbit coupling and Zeeman spin-splitting. The Majorana zero-mode becomes normalisable and exponentially localised to the vicinity of the vortex core when the superfluid is topologically non-trivial. We calculate the energy splitting due to Majorana hybridisation and identify that the $s$-wave Majorana vortices obey non-Abelian statistics. 
\end{abstract}

\maketitle

\sect{Introduction}
In two dimensions, the quasi-particle excitations of topologically ordered systems are generally non-Abelian anyons~\cite{PhysRevB.77.220501}. Possible realization of non-Abelian statistics has been studied in connection with the $\nu = 5 / 2$ fractional quantum Hall (FQH) state~\cite{PhysRevB.73.201303} and the vortex state of chiral $p_\x + \rmi p_\y $ superconductors and superfluids~\cite{PhysRevB.61.10267,Volovik1999}. Majorana fermions in the cores of superfluid vortex excitations have been actively considered in the context of $p$-wave pairing~\cite{PhysRevLett.100.027001,silaev14}, but can also occur in an $s$-wave superconductor coupled by the proximity effect to a topological insulator~\cite{Wilczek09}. Majorana vortices in the topological $s$-wave superconductor are non-Abelian, and in the same topological class~\cite{PhysRevB.77.220501} with the Moore-Read Pfaffian FQH state~\cite{MOORE1991362,doi:10.1143/PTPS.107.157}, $p_\x + \rmi p_\y $ superconductors~\cite{PhysRevLett.86.268,KITAEV20062}, and the gapped non-Abelian spin liquid phase of the Kitaev model~\cite{PhysRevLett.99.196805}. A key benefit of non-Abelian exchange statistics is that the Majorana vortices can be used to realize braiding operations. Non-commutative braiding, which amounts to exchanging two Majorana vortices adiabatically~\cite{0268-1242-27-12-124003}, is a key ingredient needed for performing quantum logic operations on a fault-tolerant topological quantum computer~\cite{RevModPhys.80.1083}.\looseness=-1

In ultra-cold atom experiments~\cite{PhysRevLett.118.123401,PhysRevLett.92.150402,Yefsah2013}, however, the densities and temperatures are low, and the scattering between fermions takes place typically in the $s$-wave channel~\cite{PhysRevLett.93.090404}. Experimentally, the ultra-cold atomic $s$-wave Fermi gas is a highly flexible quantum many-body system whose interactions, spin balance, and trapping geometries can be tuned nearly arbitrarily~\cite{Zwierlein05a,PhysRevLett.116.045304}. In the mean-field picture, it is possible to achieve band inversion leading to a topological phase~\cite{1367-2630-13-6-065004,PhysRevB.77.220501} by combining the effects of two-dimensional spin-orbit (SO) coupling~\cite{PhysRevLett.109.095302,PhysRevLett.109.095301}, and spin imbalance~\cite{Zwierlein27012006,Partridge27012006}. Recently, tunable 2D SO coupling was demonstrated experimentally~\cite{Huang16}. These advances open up a promising perspective for creating a highly-controlled atomic topological superfluid~\cite{PhysRevA.85.033622,PhysRevA.86.063604} in the laboratory in the near future. 

Despite the promising recent experimental progress, Majorana zero-modes in vortices of an $s$-wave Fermi superfluid with SO coupling and a Zeeman field remain relatively poorly understood. Here, we derive analytically the Majorana vortex zero-mode, and use it to calculate the energy splitting due to inter-vortex tunnelling. The tunnelling generally lifts the zero-mode degeneracy~\cite{PhysRevLett.103.110403}, and has been studied for the Moore-Read state~\cite{PhysRevLett.103.076801}, Kitaev's honeycomb model~\cite{LAHTINEN20082286}, and $p$-wave superconductors~\cite{PhysRevLett.103.107001,PhysRevB.82.094504,PhysRevA.82.023624}. Hybridisation of Majorana fermions in dense vortex lattices gives rise to a band structure~\cite{PhysRevB.92.134520,PhysRevB.92.134519,PhysRevB.88.064514}. Of particular interest for experimentally controlled quantum simulation are topologically non-trivial bands of Majoranas~\cite{1367-2630-13-10-105006}, and the possibility of flattening them~\cite{10.1088/1751-8121/aaf25c}.\looseness=-1

\sect{Physical system}
The Hamiltonian of a two-dimensional Fermi gas with spin-orbit coupling and spin-imbalance reads~\cite{ketterson1999superconductivity}
\begin{equation}
\label{eqn:micHmom-r}
\begin{split}
&\hat{H}(t) - \sum_\sigma \mu_\sigma \hat{N}_\sigma \\
&= \sum_{\alpha \beta} \int \d \r \, 
\hat{\psi}_\alpha^\dagger (\r, t)
\left[
\hat{K}_{\alpha \alpha}(\r) \delta_{\alpha \beta} + \hat{K}_{\alpha \beta}(\r)
\right] 
\hat{\psi}_\beta(\r,t)
\\
&\qquad + \frac{1}{2}  \sum_{\delta \gamma \alpha \beta}  \int \d \r   \int \d \rp  \\ 
&\qquad \qquad  \hat{\psi}_\delta^\dagger (\r, t)  \hat{\psi}_\gamma^\dagger (\rp, t) U_{\delta \gamma \alpha \beta}(\r,\rp) \hat{\psi}_\alpha(\rp,t)  \hat{\psi}_\beta(\r,t),
\end{split}
\end{equation}
where  $\hat{N}_\sigma =  \int \d \r \,  \hat{\psi}_\sigma^\dagger (\r, t) \hat{\psi}_\sigma(\r, t) $ is the total number of atoms with spin $\sigma = \left\lbrace \ua,\da \right\rbrace$, the operator-valued field $\hat{\psi}_{\sigma}^\dagger (\textbf{r},t)$ obeys Fermi anti-commutation relations and creates a spin-$\sigma$ fermion at location $\r$ at time $t$, 
$\hat{K}_{\sigma \sigma}(\r) = - \frac{\hbar^2\nabla^2}{2m} - \mu_\sigma$,
$\bar{\mu} = \left( \mu_\ua + \mu_\da \right) / 2$, $\zmn = \left( \mu_\ua - \mu_\da \right) / 2$ so that $\mu_\ua = \bar{\mu} + \zmn$ and $\mu_\da = \bar{\mu} - \zmn$. Here $\zmn$ is a Zeeman field creating an energy splitting between the spin components. The SO coupling is represented by $\hat{K}_{\ua \da} = -\rmi \lambda (\partial_y + \rmi \partial_x)$, $\hat{K}_{\da \ua} = -\rmi \lambda (\partial_y - \rmi \partial_x)$ with $\hat{K}_{\ua \da} = \hat{K}_{\da \ua}^\dagger $ and $\hat{K}_{\ua \da}  = -\hat{K}_{\da \ua}^*$. 

We now assume that the two-body interaction is spin-independent, $U_{\delta \gamma \alpha \beta}(\r,\rp)  = U(\r,\rp)  \delta_{\alpha \gamma} \delta_{\delta  \beta}$, and represents contact interactions, $U(\r,\rp)  = - V(\r) \delta(\r - \rp)$, where the sign convention is chosen such that $V(\r) > 0$ corresponds to attractive contact interactions.

Superfluidity in spin-balanced Fermi gases results from the formation of Cooper pairs, bound states of two fermions around the Fermi surface with opposite momenta $+\k$ and $-\k$. For spin-imbalanced pairing~\cite{PhysRevLett.120.253002}, theoretical predictions such as FFLO~\cite{PhysRev.135.A550,ZhEkspTeorFiz471136} and Sarma phases~\cite{SARMA19631029} together with deformed Fermi surface superfluidity~\cite{PhysRevLett.88.252503} have been presented. In the FFLO phase the Cooper pairs have a non-zero center-of-mass momentum which gives rise to a periodic spatial modulation of the order parameter and also possibly of density. While spin-imbalance can thus influence the pairing and subsequently the dynamics of topological defects by frustrating the Cooper pairing, in experiments the excess non-superfluid particles are typically spatially separated from the completely paired BCS superfluid~\cite{Partridge27012006,Liao2010,PhysRevLett.97.030401}. 

In typical current experimental conditions in ultra-cold atoms, the pairs form in the $s$-wave state with zero angular momentum. The Pauli principle then necessitates that the spin state be a singlet. Since the Cooper pairs have no angular momentum and no net spin, the relative pair wavefunction can be fully characterised by a single complex amplitude $\Delta(\r,t)$.

Ignoring now all the other quantum correlations apart from the pair correlations captured by the order parameter, the mean-field Hamiltonian describing quasi-particle excitations in the superfluid Fermi gas without spin-rotational invariance reads $\mathcal{H}(t) = \int \d  \r\, \begin{pmatrix} \boldsymbol{\hat{\Psi}}^\dagger (\r, t) & \boldsymbol{\hat{\Psi}}(\r, t) \end{pmatrix} H(\r,t) \begin{pmatrix} \boldsymbol{\hat{\Psi}}(\r, t) & \boldsymbol{\hat{\Psi}}^\dagger(\r, t)\end{pmatrix}^\mathrm{T}$, where $\boldsymbol{\hat{\Psi}}^\dagger (\r, t) = \begin{pmatrix} \hat{\psi}_\sigma^\dagger (\r, t) & \hat{\psi}_\sp^\dagger (\r, t) \end{pmatrix}$ is the 4-component Nambu spinor,  and 
\begin{equation}
\label{eqn:HswaveFermiSF-r-Liu}
\begin{split}
H(\r,t) &=   \begin{pmatrix}
  \hat{K}_{\ua \ua} & \hat{K}_{\ua \da} & 0 & \Delta(\r,t)  \\
\hat{K}_{\da \ua} & \hat{K}_{\da \da} & -\Delta(\r,t) & 0 \\
 0 & -\Delta^*(\r,t) & -\hat{K}_{\ua \ua}^* & \hat{K}_{\da \ua} \\
 \Delta^*(\r,t) & 0 &  \hat{K}_{\ua \da} & -\hat{K}_{\da \da}^* 
\end{pmatrix}\\
\end{split}
\end{equation}
is the Bogoliubov--De Gennes (BdG) matrix representation. In what follows, we are not interested in explicit time dependence and therefore drop the labels.

In the mean-field picture, we replace the many-body problem~\eqref{eqn:micHmom-r} with the effective single-particle Hamiltonian~\eqref{eqn:HswaveFermiSF-r-Liu} parametrised by the pair potential (order parameter) $\Delta(\r)$. The pair potential together with the chemical potential is determined by optimising the effective Hamiltonian~\eqref{eqn:HswaveFermiSF-r-Liu} such that it minimises the total free energy. The result for the pair potential is 
\begin{equation}
\label{eqn:OrderP}
\begin{split} 
\Delta(\r) &= V \langle \hat{\psi}_\ua (\r) \hat{\psi}_\da (\r) \rangle \\
&=  V \sum_\nu \left[ u_{\nu ,\ua} (\r) v^*_{\nu, \da} (\r)f_\nu \right. \\
& \qquad\qquad\qquad \left. +  u_{\nu, \da } (\r)  v^*_{\nu, \ua } (\r)   (1 - f_\nu) \right],
\end{split}
\end{equation}
and for the atomic densities
\begin{equation}
\label{eqn:ChemPot}
\begin{split} 
n_\sigma(\r) &= 
\sum_\nu \left[ \left|u_{\nu ,\sigma} (\r)\right|^2 f_\nu+  \left|v_{\nu ,\sigma} (\r)\right|^2 (1 - f_\nu) \right],
\end{split}
\end{equation}
where $f_\nu = 1/\left[\rme^{E_\nu/(k_\mathrm{B}T)} + 1\right]$ is the Fermi distribution for occupations and $T$ is the temperature. We set $T = 0$. We have expressed the order parameter in terms of the amplitudes $u$ and $v$, defined by the Bogoliubov-Valatin transformations $\hat{\psi}_\sigma (\r) = \sum_{\nu} \left[u_{\nu, \sigma}(\r) \hat{\gamma}_{\nu} + v^*_{\nu, \sigma}(\r) \hat{\gamma}_{\nu}^\dagger  \right]$, $\hat{\gamma}_\nu^\dagger = \int \d \r\, \sum_\sigma \left[ u_{\nu, \sigma}(\r) \hat{\psi}_\sigma^\dagger(\r) + v_{\nu, \sigma}(\r) \hat{\psi}_\sigma(\r)\right]$, which diagonalise the effective single-particle mean-field Hamiltonian.\looseness=-1

\sect{Symmetries of the Hamiltonian}
To obtain the static properties of Majorana fermions at the vortex cores in a topological $s$-wave Fermi superfluid, we need to solve the fundamental eigenvalue equation
\begin{equation}
\label{eqn:BdG_s-wave-matrix}
H(\r) \textbf{h}_\nu  = E_\nu \textbf{h}_\nu 
\end{equation}
for the zero-mode $\nu = 0$ with $E_0 = 0$. When considering equations of the form~\eqref{eqn:BdG_s-wave-matrix}, it is instructive to first understand the symmetries of the underlying Hamiltonian.

The BdG Hamiltonian $H(\r,t)$ possesses a particle-hole symmetry (PHS) defined by $\lbrace \Xi_\mathrm{s-wave},H(\r,t)\rbrace = 0$ with $\Xi_\mathrm{s-wave} = \rme^{\rmi \theta}\tau_\x \otimes 1_\sigma K$, where $K$ is the complex conjugation operator (in momentum space, $K$ additionally flips the sign of the momentum), $\tau_\y$ ($\sigma_\z$) is the Pauli matrix in particle-hole (spin) space, and $\theta \in \field{R}$. As a result of the particle-hole symmetry, if $\textbf{h}_\nu = \begin{pmatrix} u_{\nu,\uparrow} & u_{\nu,\downarrow} & v_{\nu,\uparrow} & v_{\nu,\downarrow} \end{pmatrix} ^\mathrm{T}$ is a solution of Eq.~\eqref{eqn:BdG_s-wave-matrix} with energy $E_\nu$, then $\Xi_\mathrm{s-wave} \textbf{h}_\nu = \rmi\, \rme^{\rmi \theta} \begin{pmatrix} -v_{\nu,\uparrow}^* & v_{\nu,\downarrow}^* & u_{\nu,\uparrow}^* & -u_{\nu,\downarrow}^* \end{pmatrix}^\mathrm{T}$ is a solution with energy $-E_\nu$. 

Additionally, if and only if $\Delta(\r,t) \in \mathbb{R}$ and $\zmn = 0$, the BdG Hamiltonian possesses a time-reversal symmetry defined by $\left[ T, H(\r,t) \right] = 0$, where $T = \rme^{\rmi \theta} 1_\tau \otimes \sigma_\y K$ ($T^2 = -1$, for half-integer spin), and $\theta$ is arbitrary. We have defined $1_\tau$ to be the identity matrix in the particle-hole space, while $\sigma_\y$ is the Pauli matrix in spin space. 

A straightforward calculation shows that non-degenerate zero-modes (that is, the modes $H(\r) \textbf{h}_0  =  E_0 \textbf{h}_0 $ with $E_0 = 0$), must always satisfy the important symmetry
\begin{equation}
\label{eqn:Majorana_swave_rel1-Liu}
u_{0,\sigma} = \rme^{-\rmi \theta}v^*_{0,\sigma},
\end{equation}
which is the so-called Majorana property making the zero-mode special. The Majorana property ensures that the quasi-particle operators
\begin{equation}
\label{eqn:Majorana_swave_rel1-MF}
\hat{\gamma}_0^\dagger = \int \d \r\, \sum_\sigma \left[ u_{0, \sigma}(\r) \hat{\psi}_\sigma^\dagger(\r) + v_{0, \sigma}(\r) \hat{\psi}_\sigma(\r)\right]
\end{equation}
satisfy $\hat{\gamma}_0^\dagger = \hat{\gamma}_0$; they represent Majorana fermions.

Let us temporarily assume that time-reversal symmetry is respected, i.e. $\Delta(\r) \in \mathbb{R}$ and $\zmn= 0$. If $\textbf{h}_\nu = \begin{pmatrix} u_{\nu,\uparrow} & u_{\nu,\downarrow} & v_{\nu,\uparrow} & v_{\nu,\downarrow} \end{pmatrix} ^\mathrm{T}$ is a solution of Eq.~\eqref{eqn:BdG_s-wave-matrix} with energy $E_\nu$, then $T \textbf{h}_\nu$ is a solution with the same energy $E_\nu$. This is the pair of the zero-energy Majorana mode. With time-reversal symmetry, the Majorana probability densities are identically overlapping in space: $|\textbf{h}_0|^2 = |T\textbf{h}_0|^2 = 2|u_{0,\uparrow}|^2 + 2 |u_{0,\downarrow}|^2$. We need to break the time-reversal symmetry to separate them spatially, for example, by having a vortex in the order parameter.

A broken time-reversal symmetry $T$, a broken spin-rotational symmetry (due to spin-orbit coupling), and an unbroken PHS $\Xi_\mathrm{s-wave}$ result in the BdG Hamiltonian $H(\r)$ belonging to the symmetry class D in the Altland-Zirnbauer classification~\cite{PhysRevB.55.1142}. According to Ref.~\cite{PhysRevB.82.115120}, point defects (such as the vortices in the order parameter considered here) with a Hamiltonian that belongs to the class D are associated with a $\mathbb{Z}_2$-valued topological invariant. This means that in the topologically non-trivial regime the number of exact Majorana zero-modes is given by $W \text{ mod }2$, where $W$ is the sum over all vortex winding numbers, with the rest hybridising in pairs forming a band structure. 
Generally, when multiple vortices are brought close together, the Majorana zero-modes hybridize into a band structure, leading to complex fermion states at positive and negative energy~\citep{PhysRevB.92.134520,PhysRevB.92.134519,PhysRevB.88.064514}. We calculate the energy splitting in Eq.~\eqref{eqn:HoppingAmplitude}. For periodic vortex lattices, such as the square and the triangular lattice, the resulting low-energy theory is typically gapped and topologically non-trivial~\cite{PhysRevB.92.134519}.

\subsect{Eigenvalue equation for the zero-mode}
We now include a vortex in the order parameter,
\begin{equation}
\label{eqn:OrderP-vortex}
\Delta  (\r) = \OP (r) \rme^{\rmi \ell \varphi },
\end{equation}
where $r$ and $\varphi$ are polar coordinates centred on the vortex. Energy considerations require that $\Delta  (\textbf{0}) = 0$ at the vortex core. The integer $\ell$ denotes the vorticity, and $\OP(r)$ is a real function of $r$ that vanishes at $r = 0$. In what follows, we will solve Eq.~\eqref{eqn:BdG_s-wave-matrix} analytically for the zero-mode $\textbf{h}_0$ that exists in the core of the vortex~\eqref{eqn:OrderP-vortex}.

For convenience, let us first apply the unitary transformation $\mathcal{U_\mathrm{s}}$, which transforms the basis as $\mathcal{U_\mathrm{s}}\begin{pmatrix}
u_{0,\uparrow} &
u_{0,\downarrow} &
v_{0,\uparrow} &
v_{0,\downarrow}
\end{pmatrix}^\mathrm{T} = \begin{pmatrix}
u_{0,\downarrow} &
v_{0,\downarrow} &
v_{0,\uparrow} &
u_{0,\uparrow} 
\end{pmatrix}^\mathrm{T} $, and shuffles the zeros in the rows and columns of the Hamiltonian $H(\textbf{r})$ [Eq.~\eqref{eqn:HswaveFermiSF-r-Liu}] to give
\begin{equation}
\label{eqn:Hamshuffled}
\begin{split}
H^{(\mathcal{U_\mathrm{s}})}(\r) \equiv \mathcal{U_\mathrm{s}}H(\r)\mathcal{U_\mathrm{s}}^{-1} 
&= \begin{pmatrix}
D_\downarrow & M \\
M^\dagger & -D_\uparrow
\end{pmatrix},
\end{split}
\end{equation}
where $D_\sigma = \text{diag}( \hat{K}_{\sigma \sigma}, -\hat{K}_{\sigma \sigma})$ and 
\begin{equation}
\label{eqn:HammatM}
M = \begin{pmatrix}
-\Delta(\r)  & -\rmi \lambda (\partial_y - \rmi \partial_x) \\
-\rmi \lambda (\partial_y + \rmi \partial_x) & \Delta^*(\r) \\
\end{pmatrix}.
\end{equation}

Setting $D_\sigma = 0$ decouples the spins, and the Hamiltonian reduces to a $2\times 2$ BdG matrix corresponding to the Fu-Kane model at neutrality, whose Majorana modes have been considered~\cite{PhysRevB.91.165402}. This is the low-energy regime near $\mu_\sigma = 0$, which is the topological transition point in typical $p$-wave systems. The Majorana zero-modes have also been considered for the Fu-Kane model at $\mu \neq 0$~\cite{PhysRevB.82.094504}.

However, here we consider the full $4 \times 4$ structure of Eq.~\eqref{eqn:BdG_s-wave-matrix}:
\begin{subequations}
\label{eqn:BdG_s-wave-sys}
\begin{align}
\label{eqn:BdG_s-wave-sys-a}
\hat{K}_{\ua \ua} u_{\nu,\ua} + \hat{K}_{\ua \da} u_{\nu,\da} + \Delta (\r) v_{\nu,\da} &= E_\nu u_{\nu,\ua}, \\
\label{eqn:BdG_s-wave-sys-b}
-\hat{K}_{\ua \ua} v_{\nu,\ua} +\hat{K}_{\da \ua} v_{\nu,\da} - \Delta^* (\r) u_{\nu,\da} &= E_\nu v_{\nu,\ua}, \\
\label{eqn:BdG_s-wave-sys-c}
\hat{K}_{\da \ua} u_{\nu,\ua} + \hat{K}_{\da \da} u_{\nu,\da} - \Delta (\r) v_{\nu,\ua} &= E_\nu u_{\nu,\da}, \\
\label{eqn:BdG_s-wave-sys-d}
\hat{K}_{\ua \da} v_{\nu,\ua} -\hat{K}_{\da \da}  v_{\nu,\da} + \Delta^*(\r)  u_{\nu,\ua} &= E_\nu v_{\nu,\da}. 
\end{align}
\end{subequations}
When looking for zero-modes, the system~\eqref{eqn:BdG_s-wave-sys} can be simplified by using the symmetry property~\eqref{eqn:Majorana_swave_rel1-Liu}, that is, we seek simultaneous eigenstates of the PHS. There are two possibilities for $\theta$, namely $\theta=\theta_0, \theta_0 + \pi$, which both satisfy the Majorana relation. This means that with the final solution we must allow for both possibilities. In both cases, we obtain 
\begin{subequations}
\label{eqn:BdG_s-wave-1-sys}
\begin{align}
\label{eqn:BdG_s-wave-1-sys-a}
\hat{K}_{\ua \ua} u_{0,\ua} + \hat{K}_{\ua \da} u_{0,\da} + \Delta(\r) u^*_{0,\da} &= 0, \\
\label{eqn:BdG_s-wave-1-sys-c}
\hat{K}_{\da \ua} u_{0,\ua} + \hat{K}_{\da \da} u_{0,\da} - \Delta(\r) u^*_{0,\ua} &= 0,
\end{align}
\end{subequations}
a coupled $2 \times 2$ system, which must be solved for $u_{0,\ua}$ and $u_{0,\da}$.

In polar coordinates, the spin-orbit coupling terms read $\hat{K}_{\ua \da}(\r) = \lambda\rme^{-\rmi  \varphi}\left[\partial_r - (\rmi /r)\partial_\varphi \right]$. Observing the azimuthal symmetry, let us try a separable ansatz with angular momentum eigenstates,
\begin{equation}
\label{eqn:BdG_s-wave-ansatz}
u_{0,\sigma} = \rme^{\rmi m_\sigma \varphi} F_\sigma(r)
\rme^{- \frac{\OP}{\lambda} r},
\end{equation}
with the assumption that $\OP$ has no $r$-dependence. Physically, this approximation treats the vortices as points with only a phase profile. Substitution into the system~\eqref{eqn:BdG_s-wave-1-sys} shows that we can eliminate the angular dependence by taking $m_+ = \zeta$ and $m_- = -\ell$, where $m_\pm \equiv \pm m_\sp - m_\sigma $,  $\zeta = \pm 1$ with $\zeta =+ 1$ for $\sigma = \ua$ and $\zeta =- 1$ for $\sigma = \da$, and $\sigma^\prime= \ua, \da$ if $\sigma = \da, \ua$ respectively. This implies $m_\sp =( \ell + \zeta)/2 $ and  $m_\sigma = (\ell- \zeta )/2$. The only requirement here is that $\ell \in \field{Z}$. Explicitly, the coupled system~\eqref{eqn:BdG_s-wave-1-sys} then reads
\begin{subequations}
\label{eqn:BdG_s-wave-4-simplf-aux-sys}
\begin{align}
\label{eqn:BdG_s-wave-4-simplf-aux-sys-a}
- F_\ua^{\prime \prime}
&- \left(\frac{1}{r} - 2\frac{\OP }{\lambda} \right)  F_\ua^\prime   
+ \left( -\frac{\OP^2}{\lambda^2} +  \frac{\OP }{\lambda r}  \right)  F_\ua   \\
\notag &
+ \frac{m_\ua^2 F_\ua}{r^2} 
+ \frac{2m \lambda}{\hbar^2} \left( F_\da^\prime  + \frac{m_\da F_\da}{r} \right) 
   = \frac{2m} {\hbar^2} \mu_\ua F_\ua, \\
  \label{eqn:BdG_s-wave-4-simplf-aux-sys-b}
 -F_\da^{\prime \prime}
 &-\left( \frac{1}{r} - 2 \frac{\OP}{\lambda}\right)  F_\da^\prime   
 + \left(- \frac{\OP^2 }{\lambda^2} + \frac{ \OP }{\lambda r} \right) F_\da   \\
 \notag &
 + \frac{m_\da^2 F_\da}{r^2}
+ \frac{2m \lambda}{\hbar^2}  \left(- F_\ua^\prime + \frac{m_\ua F_\ua}{r} \right)   
= \frac{2m} {\hbar^2} \mu_\da F_\da.
\end{align}
\end{subequations}
Below, we solve analytically the coupled system~\eqref{eqn:BdG_s-wave-4-simplf-aux-sys} for $F_\ua(r)$ and $F_\da(r)$.

\subsect{Series solution for $F_\sigma(r)$}
The vortex core ($r = 0$) is an irregular singular point of Eq.~\eqref{eqn:BdG_s-wave-4-simplf-aux-sys} where numerical methods are inherently unstable. An obvious solution is to start the numerical integration at a point $r_0 > 0$ thus avoiding the singularity, but in this case we need to know accurately the initial condition at the rather arbitrary point $r_0$. It is therefore highly desirable to have analytic solutions, at least around the vortex core. In what follows we solve the system~\eqref{eqn:BdG_s-wave-4-simplf-aux-sys} using the Wasow method~\cite{wasow2002asymptotic}, a generalisation of the Fr\"obenius series method for coupled systems of differential equations with irregular singular points.

The details for solving the system~\eqref{eqn:BdG_s-wave-4-simplf-aux-sys} are shown in Appendix~\ref{Sec:AppA}. For definitess, we set $\ell = +1$. The result is a series solution for $F_\sigma(r)$ around the origin that can be evaluated analytically to arbitrary order, and reads
\begin{subequations}
\label{eqn:BdG_s-wave-SOLUTION}
\begin{align}
 F_\ua (r) &= 
F_\ua(0)
+\frac{\OP F_\ua(0)}{\lambda }r  \\
& \notag + \left\lbrace \frac{\lambda  m}{\hbar ^2}F_\da^\prime(0)
+ \left[ \frac{\OP^2 }{2 \lambda ^2}-\frac{ (\mu + h) m}{2 \hbar ^2}
\right]F_\ua(0)
\right\rbrace r^2 \\
\notag & +  \left\lbrace \left[\frac{\OP^3 }{6 \lambda ^3}-\frac{\OP  (\mu+h)  m}{2 \lambda  \hbar ^2}-\frac{4 \OP  \lambda  m^2}{9 \hbar ^4}\right] F_\ua(0) \right. \\
\notag & \left. \qquad +\frac{11 \OP m}{9 \hbar ^2}F_\da^\prime(0) \right\rbrace r^3
+ \mathcal{O}(r^4) ,\\
F_\da (r) &= F_\da^\prime(0) r +\OP\left(\frac{F_\da^\prime(0)}{\lambda }-\frac{2 F_\ua(0) m}{3 \hbar ^2}\right) r^2  \\
\notag &
+\left\lbrace \frac{2 \OP^2 \hbar ^4+\lambda ^2 m \hbar ^2 (h-\mu )-2 \lambda ^4 m^2}{4 \lambda ^2 \hbar ^4} F_\da^\prime(0)  \right. \\
&\notag \left.
+ \frac{  -8 \OP^2 \hbar ^2m + 3 h \lambda ^2 m^2+3 \lambda ^2 \mu  m^2}{12 \lambda  \hbar ^4} F_\ua(0) \right\rbrace r^3 \\
&\notag
+ \mathcal{O}(r^4).
\end{align}
\end{subequations}
While $F_\ua(0)$ can be finite, the boundary condition $F_\da(0) = 0$ is forced to keep $F_\da$ finite at the origin, which results from the series solution for $F_\da$ starting with the power $r^{-1}$ whereas that for $F_\ua$ starts with $r^0$. That one spin component is zero and the other finite at the vortex core has been observed in all numerical studies of the Majorana zero-mode in $s$-wave Fermi gases~\cite{PhysRevA.85.021603,PhysRevA.85.013622}. In principle the system~\eqref{eqn:BdG_s-wave-4-simplf-aux-sys} needs four initial conditions, but requiring that $F_\da$ be finite at the origin consumes two of them leaving only the two initial conditions $F_\ua(0)$ and $F^\prime_\da(0)$ in the solution~\eqref{eqn:BdG_s-wave-SOLUTION}.

\subsect{Existence of the Majorana zero-mode}
The existence of the analytical series solutions~\eqref{eqn:BdG_s-wave-SOLUTION} alone does not guarantee the existence of a topologically protected Majorana zero-mode. The condition for the physical existence of the zero-mode is that it remain normalisable as $r \to \infty$.
 
In the limit $r \to \infty$, the kinetic energy and terms in $1/r$ can be dropped in system~\eqref{eqn:BdG_s-wave-4-simplf-aux-sys} to obtain
\begin{subequations}
\label{eqn:BdG_s-wave-5-sys}
\begin{align}
\label{eqn:BdG_s-wave-5-sys-a}
\lambda  F_\da^\prime(r)+\OP F_\da(r)-  F_\ua(r) (\bar{\mu}+ \zmn )&=0, \\
\label{eqn:BdG_s-wave-5-sys-c}
\lambda  F_\ua^\prime(r)+\OP F_\ua(r)+  F_\da(r) (\bar{\mu}- \zmn )&=0.
\end{align}
\end{subequations}
Physically, we are interested in only the solutions for system~\eqref{eqn:BdG_s-wave-5-sys} that are normalisable. For example, if $\lambda > 0$ and $\OP < 0$, the only normalisable asymptotic solution is
\begin{equation}
\label{eqn:BdG_s-wave-6}
F_\ua(r) = C\, \textrm{exp}{\left[-\frac{\OP+\sqrt{h^2-\bar{\mu} ^2}}{\lambda } r \right]} = \frac{\bar{\mu} - \zmn    }{\sqrt{\zmn^2-\bar{\mu} ^2} } F_\da(r),
\end{equation}
where $C$ is a constant. Even then, the solution~\eqref{eqn:BdG_s-wave-6} is normalisable only when $\zmn >\sqrt{\bar{\mu}^2 + |\Delta|^2}$. The value at equality corresponds to the critical Zeeman field marking the phase transition to the topological regime with a uniform order parameter, $\Delta(\r) = \Delta$~\cite{10.21468/SciPostPhys.5.2.016}, which here is associated with the boundedness of the Majorana zero-mode at infinity. However, it is not straightforward to find initial conditions $F_\ua(0)$ and $F^\prime_\da(0)$ such that the zero-mode has the desired large-$r$ asymptotics.

In principle, $\bar{\mu}$ and $\Delta$ must be obtained self-consistently in conjunction with solving the BdG eigenvalue equation. Considering a uniform vortex-free system, we can solve Eqs.~\eqref{eqn:OrderP},~\eqref{eqn:ChemPot}, and~\eqref{eqn:BdG_s-wave-sys} self-consistently to find that the superfluid is topologically non-trivial with the parameters $\lambda= 5$, $h= 10$, $\OP = -1.587$, $\bar{\mu} = -5.987$. Here all units are measured with respect to $\hbar = m =  1$. We have used the pair binding energy of $E_\mathrm{b} = 0.05$ in the renormalisation of the gap equation, which otherwise diverges logarithmically. The pair binding energy determines the state of the Cooper pairs across the BCS-BEC crossover~\cite{10.21468/SciPostPhys.5.2.016}. From Eq.~\eqref{eqn:BdG_s-wave-6}, therefore, we know that with these parameters there is an exponentially decaying asymptotic solution, and we have performed a direct search for $F_\ua(0)$ and $F^\prime_\da(0)$ that match with this asymptotic behaviour. 

\begin{figure}[t]
  \centering
        \begin{tikzpicture}
        \def\x{4.5};        \def\y{3.0};
        \def\v{-1.1};   \def\vv{-1.3};;    \def\w{10.8};	     \def\u{10.8};
			 \node at (0.5*\x,1.5*\y) {    \includegraphics[width=0.43\textwidth,angle=0]{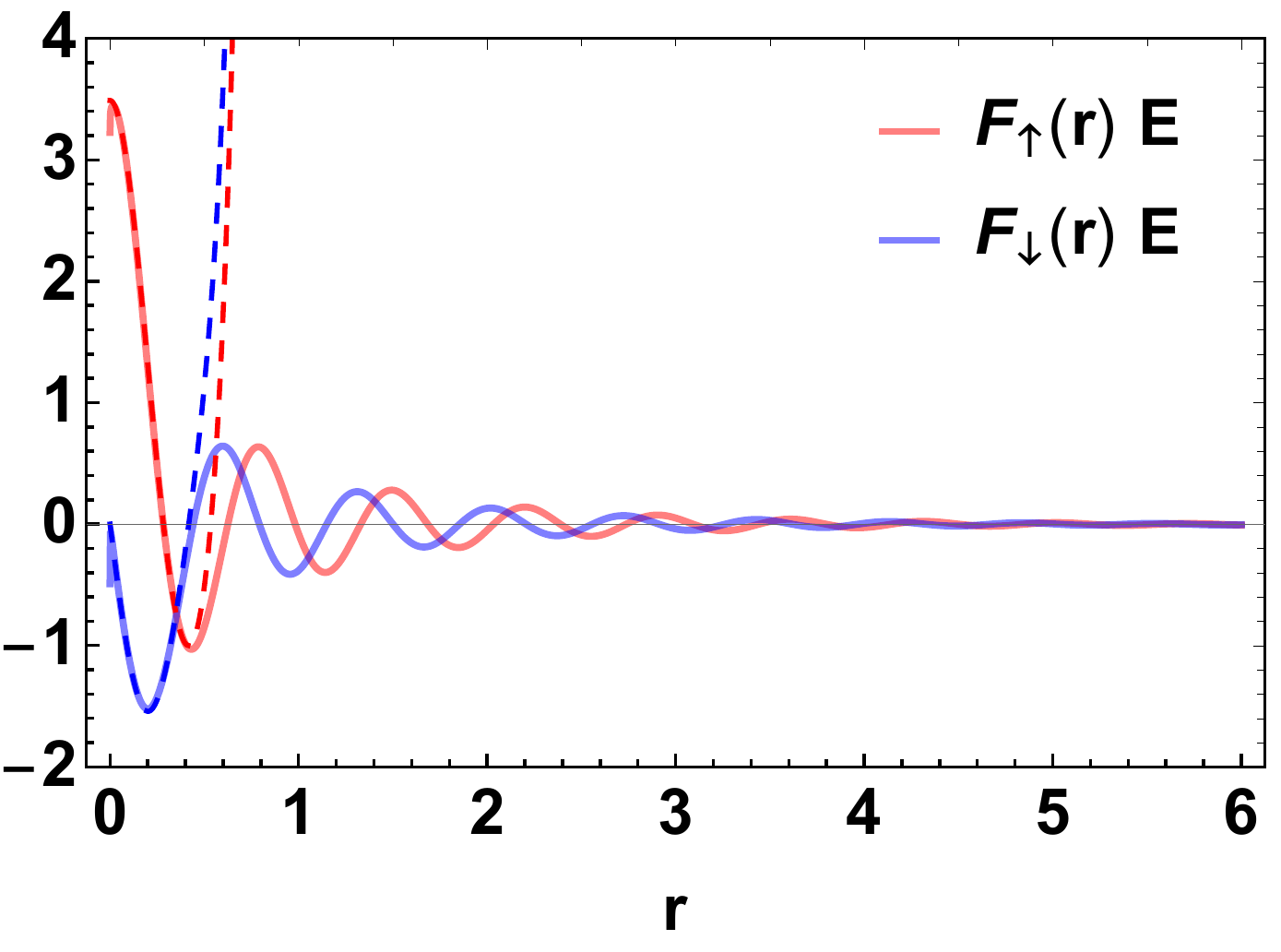}};
                  
      \end{tikzpicture}
      
      \caption{ \label{fig:1} 
      Majorana zero-mode housed by a vortex with unit circulation in an $s$-wave Fermi gas. The vortex core corresponds to $r = 0$. The solid lines correspond to numerical integration of the system~\eqref{eqn:BdG_s-wave-4-simplf-aux-sys}, and the dashed lines correspond to the series solution~\eqref{eqn:BdG_s-wave-SOLUTION} upto $\mathcal{O}(r^{10})$. Here $E \equiv \mathrm{exp}\left(-\frac{\OP}{\lambda} r \right)$. The initial condition for the numerical evaluation at $r = r_0$ is provided by the analytical solution~\eqref{eqn:BdG_s-wave-SOLUTION}. The zero-mode solution becomes bounded at $r \to \infty$ when the superfluid is topologically non-trivial. The parameters are found self-consistently for the uniform vortex-free state, and read $\lambda= 5$, $h= 10$, $\OP = -1.587$, $\bar{\mu} = -5.987$, and $E_\mathrm{b} = 0.05$. Within numerical accuracy the Majorana zero-mode is normalisable for $3.488885 < F_\ua(0) < 3.488886$ and $F^\prime_\da(0) = -12.518$. We have set $\hbar = m = 1$.      
      }   
\end{figure}

We retain terms upto $\mathcal{O}(r^{10})$, and use the analytical solution~\eqref{eqn:BdG_s-wave-SOLUTION} to evaluate accurate initial conditions for a numerical integration of the system~\eqref{eqn:BdG_s-wave-4-simplf-aux-sys}. The truncated analytical series solution agrees with the numerical integration for $r \lesssim 0.5$ (Fig.~\ref{fig:1}). The analytic series expansion can be developed to arbitrary precision. As expected, the numerical solution becomes unstable as the vortex core is approached, but more importantly it can be used to study the boundedness of the mode as $r \to \infty$. We find a bounded zero-mode in vortices with unit positive circulation with the specific parameter values above for $3.488885 < F_\ua(0) < 3.488886$ and $F^\prime_\da(0) = -12.518$. The full solution including azimuthal dependence is then given by Eq.~\eqref{eqn:BdG_s-wave-ansatz}.

\sect{Majorana energy splitting}
We now use the Majorana zero-mode to calculate the energy splitting in an $s$-wave Fermi gas that results from hybridisation between the Majorana modes of two vortices located at points $\r_1$ and $\r_2$. Then, $\Delta  (\r) \approx \Delta_i  (\r) $ near vortex $i$ ($i = 1,2$), where $\Delta_i  (\r)$ is the single-vortex pair function~\eqref{eqn:OrderP-vortex} for vortex $i$. Introduction of a single-vortex background phase $\theta_i$, $\Delta_i(\r) = \OP \rme^{\rmi  (\ell_i\varphi_i + \theta_i)}$ with $\varphi_i$ the azimuthal angle measured with respect to vortex $i$, amounts to the Majorana mode changing by $u_{0,\sigma}(\r-\r_i) \to \rme^{\rmi \frac{\theta_i}{2}  }u_{0,\sigma}(\r-\r_i)$, leaving invariant the system~\eqref{eqn:BdG_s-wave-1-sys}.

In the tight-binding approximation, the Hamiltonian $H_\Delta$ describing hopping between the two vortices, with a straightforward generalisation to lattices of Majorana vortices, is given by $H_\Delta = \rmi\, t_{12}\, \hat{\gamma}_{0,1} \hat{\gamma}_{0,2}$, where $\hat{\gamma}_{0,i}$ is the Majorana mode~\eqref{eqn:Majorana_swave_rel1-MF} at vortex $i$, and $ t_{12}  = \langle \textbf{g}_0^{(2)} | H^{(\mathcal{U_\mathrm{s}})}  |  \textbf{g}_0^{(1)} \rangle 
 = \langle \textbf{g}_0^{(2)} \left|  \begin{pmatrix}
D_\downarrow & M \\
M^\dagger & -D_\uparrow
\end{pmatrix}  \right|  \textbf{g}_0^{(1)} \rangle$ is the overlap integral that gives the energy splitting. Here $\textbf{g}_0^{(i)}(\r) = \begin{pmatrix}
u_{0,\downarrow}(\r-\r_i) &
u^*_{0,\downarrow}(\r-\r_i)  &
u^*_{0,\uparrow}(\r-\r_i)  &
u_{0,\uparrow}(\r-\r_i) 
\end{pmatrix}^\mathrm{T}
$
is the Majorana zero-mode centered at vortex $i$. 

For convenience, we define $M\a \equiv M - \tilde{M}$ such that $M\a$ coincides with Eq.~\eqref{eqn:HammatM} when the order parameter is given by just a single vortex at position $\r_1$. It follows that near vortex 1 $ \tilde{M}  \approx 0$, and the dominant contribution to the overlap integral comes from the vicinity of vortex 2. 
Taking $\ell_1 = \ell_2 = 1$, using Eq.~\eqref{eqn:BdG_s-wave-ansatz}, and introducing the definitions $G(\r) \equiv -\OP\,  \mathrm{exp}\left[-\frac{f^2}{\lambda} \left( |\r-\r_2|+ |\r-\r_1|\right) \right]$, $\Theta_{\sigma \sp}(\r) \equiv F_\sigma(|\r-\r_2|) F_\sp(|\r-\r_1|)$, by definition of the zero-mode at vortex 1 we obtain 
\begin{equation}
\label{eqn:HoppingAmplitude}
\begin{split}
 t_{12}  
&= 
 4 \cos{\left( \frac{\theta_1 - \theta_2}{2}    \right)}
 \int \d  \r\,  G(\r) \sin^2{\left( \frac{\varphi_1- \varphi_2}{2} \right)} \left\lbrace
 \right.  \\
 & \left. \qquad \qquad  \Theta_{\da \ua}(\r)  -\,  \Theta_{\ua \da}(\r)  
\right\rbrace .
 \end{split} 
\end{equation}
Compared with $p$-wave pairing~\cite{PhysRevLett.111.136401}, under Majorana exchange $t_{12}$ is now symmetric with respect to $\theta_1 - \theta_2$, but still anti-symmetric due to the spin degree of freedom.

It was pointed out by Fujimoto~\cite{PhysRevB.77.220501} that the existence of the Majorana zero-mode with SO coupling does not automatically guarantee their non-Abelian statistics. Majorana interchange $1 \leftrightarrow 2$ gives $t_{12} = -t_{21}$, and the U(1) gauge transformation $\theta_i$ of the Majorana fermion has the important property that when $\theta_i$ changes from $0 \to 2\pi$, the Majorana changes sign, $\hat{\gamma}_{0,i} \to -\hat{\gamma}_{0,i}$. Braiding of vortices $i$ and $j$ changes the superfluid phase at one vortex by $2\pi$ amounting to $\hat{\gamma}_{0,i}  \to \hat{\gamma}_{0,j}$, $\hat{\gamma}_{0,j}  \to -\hat{\gamma}_{0,i}$, and it was shown by Ivanov~\cite{PhysRevLett.86.268} that this property together with quantisation of $\ell_i$ gives rise to non-Abelian statistics.\looseness=-1

\sect{\label{sec:dc} Conclusions}
We have derived analytically the Majorana zero-mode in a topological Fermi superfluid with $s$-wave pairing, two-dimensional spin-orbit coupling, and a Zeeman field. We find an exponentially localised Majorana zero-mode at the vortex core only in the topologically non-trivial regime of the superfluid. We find that in the $s$-wave Fermi superfluid the Majorana fermions obey non-Abelian exchange statistics, and the energy splitting due to Majorana hybridisation is determined by both spin sectors. Knowing the Majorana zero-mode analytically paves the way for studies of quantum many-body correlations and simulation of topological quantum matter in lattices of Majorana vortices in a new experimentally well-controlled setup.

\acknowledgements
I would like to thank Andreas L\"auchli for fruitful discussions. This work was supported by the Austrian Academy of Sciences (P7050-029-011).

\bibliographystyle{apsrev4-1}
\bibliography{Majorana,Helicity_TF}

\appendix

\section{\label{Sec:AppA}Solution for the Majorana zero-mode}
\subsection{Formulation as a first-order system}

Introducing the first derivatives $x = F_\ua^\prime$ and $y = F_\da^\prime$ as auxiliary variables, we can trivially regroup the system~\eqref{eqn:BdG_s-wave-4-simplf-aux-sys} for $x^\prime$ and $y^\prime$ in terms of $x,y,F_\ua,F_\da$. This gives the equivalent matrix equation
\begin{equation}
\label{eqn:BdG_s-wave-t1-0}
\u^\prime(r) = \mathcal{M}(r) \u(r),
\end{equation}
where 
\begin{subequations}
\label{eqn:BdG_s-wave-t1-1-sys}
\begin{align}
\label{eqn:BdG_s-wave-t1-1-sys-a}
\u(r) &= \begin{pmatrix}
F_\ua \\
x \\
F_\da \\
y
\end{pmatrix}, \\
\label{eqn:BdG_s-wave-t1-1-sys-b}
 \mathcal{M}(r) &=
 \begin{pmatrix}
0 & 1 & 0 & 0\\
\Gamma_\ua  & - \left(\frac{1}{r} - 2\frac{f^2 }{\lambda} \right)  &  \frac{2m \lambda}{\hbar^2}  \frac{m_\da }{r} &  \frac{2m \lambda}{\hbar^2} \\
0 & 0 & 0 & 1\\
 \frac{2m \lambda}{\hbar^2}  \frac{m_\ua }{r} & - \frac{2m \lambda}{\hbar^2} & \Gamma_\da  & -\left( \frac{1}{r} - 2 \frac{f^2 }{\lambda}\right) 
\end{pmatrix},
\end{align}
\end{subequations}
where $\Gamma_\sigma  \equiv  -\frac{f^4}{\lambda^2} +  \frac{f^2 }{\lambda r}+ \frac{m_\sigma^2 }{r^2} - \frac{2m} {\hbar^2} \mu_\sigma$. This is a set of linear equations because $\mathcal{M}(r)$ does not depend on the components of $\u(r)$, but the coefficient matrix $\mathcal{M}(r)$ is non-constant depending on $r$. 
 
The point $r=0$ is an irregular singular point of Eq.~\eqref{eqn:BdG_s-wave-t1-0} because the coefficient matrix $\mathcal{M}(r)$ has a pole of order 2 at the origin.  

Generally, if $\mathcal{M}(r)\mathcal{M}(r^\prime) = \mathcal{M}(r^\prime)\mathcal{M}(r)\,\, \forall r,r^\prime$, then Eq.~\eqref{eqn:BdG_s-wave-t1-0} can be easly solved in terms of the matrix exponential
\begin{equation}
\label{eqn:BdG_s-wave-t1-2}
\u(r) =\u(r_0)\, \rme^{\int_{r_0}^r \mathcal{M}(s) \d s},
\end{equation}
where $\u(r_0)$ is a 4-component constant vector. However, the commutation property does not hold here. There is no general closed-form solution for differential equations of the form of Eq.~\eqref{eqn:BdG_s-wave-t1-0} where $\mathcal{M}(r)$ does not satisfy the commutation property. The Magnus series provides systematically an exact solution in terms of an infinite series of nested commutators:
\begin{equation}
\label{eqn:BdG_s-wave-t1-3}
\u(r) =\u(r_0)\, \rme^{\sum_{k = 1}^\infty  \Omega_k(s)},
\end{equation}
where
\begin{subequations}
 \begin{align}
  \Omega_1(s) &= \int_{r_0}^r \mathcal{M}(s_1)\,\d s_1, \\
  \Omega_2(s) &= \frac{1}{2}\int_{r_0}^r \,\d s_1 \int_{r_0}^{s_1} \,\d s_2\ \left[  \mathcal{M}(s_1),\mathcal{M}(s_2)\right], \\
                      \Omega_3(s) &= \ldots ,
 \end{align}
\end{subequations}
but this approach suffers from the irregular singular point at the origin as well. 

Near $r = 0$, we can always write 
\begin{equation}
\label{eqn:BdG_s-wave-t1-4}
\u^\prime(r) = \left(\frac{1}{r^g} \sum_{\nu = 0}^\infty \mathcal{M}_\nu r^\nu \right) \u(r),
\end{equation}
where $\mathcal{M}_\nu$ are constant $4 \times 4$ matrices holomorphic at $r = 0$ for which the series converges component-wise in a neighbourhood of the origin. Here $g = 2$, and $\mathcal{M}_\nu \neq 0$ for only $\nu = 0,1,2$.

\subsection{Solution for the eigenvalue equation~\eqref{eqn:BdG_s-wave-4-simplf-aux-sys} }

Generalising the vector $\u$ into the matrix $X$, and transforming $x = 1/r$ changes Eq.~\eqref{eqn:BdG_s-wave-t1-4} into
\begin{equation}
\label{eqn:BdG_s-wave-z-1}
x^{-q}X^\prime(x) = -\mathcal{M}(x) X(x),
\end{equation}
where $q = g - 2$ and 
\begin{equation}
\label{eqn:BdG_s-wave-z-1ab}
\mathcal{M}(x) = \sum_{\nu = 0}^\infty \mathcal{M}_\nu x^{-\nu}
\end{equation}
with the only non-zero matrices being
\begin{subequations}
\label{eqn:BdG_s-wave-Turritin-3-sys}
 \begin{align}
 \label{eqn:BdG_s-wave-Turritin-3-a}
  \mathcal{M}_0 &=  \begin{pmatrix}
0 & 0 & 0 & 0\\
m_\ua^2 &0  &  0& 0 \\
0 & 0 & 0 & 0\\
0&0 & m_\da^2  & 0
\end{pmatrix} ,\\
 \label{eqn:BdG_s-wave-Turritin-3-b}
\mathcal{M}_1 &=  \begin{pmatrix}
0 & 0 & 0 & 0\\
\frac{f^2 }{\lambda} & - 1 &  \frac{2m \lambda}{\hbar^2} m_\da  &0\\
0 & 0 & 0 & 0\\
 \frac{2m \lambda}{\hbar^2} m_\ua  & 0 &\frac{f^2 }{\lambda }& -1 ,\\
 \end{pmatrix} ,\\
 \label{eqn:BdG_s-wave-Turritin-3-c}
\mathcal{M}_2 &=  \begin{pmatrix}
0 & 1 & 0 & 0\\
-\frac{f^4}{\lambda^2} - \frac{2m} {\hbar^2} \mu_\ua & 2\frac{f^2 }{\lambda}   & 0&  \frac{2m \lambda}{\hbar^2} \\
0 & 0 & 0 & 1\\
0& - \frac{2m \lambda}{\hbar^2} & -\frac{f^4}{\lambda^2} - \frac{2m} {\hbar^2} \mu_\da  & 2 \frac{f^2 }{\lambda}  ,\\
\end{pmatrix}.
 \end{align}
\end{subequations}

The matrix $\mathcal{M}(x)$ is holomorphic at $x = \infty$ meaning that there exists a convergent expansion of the form~\eqref{eqn:BdG_s-wave-z-1ab} for sufficiently large $x_0$ such that $|x| > x_0$. If the integer $q + 1 > 0$, the singular point is irregular, and if $q + 1 = 0$, the singular point is regular. For us $q = 0$. Our goal is to reduce Eq.~\eqref{eqn:BdG_s-wave-z-1} (where $q = 0$) through formal transformations into a system with $q = -1$, which corresponds to a regular singular point at the origin $x = \infty$, and can be solved in terms of more standard Fr\"obenius series ansatz methods.

To form our starting point, we transform $\mathcal{M}_0$ into the Jordan canonical form (JCF) (we define the JCF as having the 1's on the superdiagonal). The matrix $\mathcal{M}_0$ is brought to the JCF by the non-singular constant similarity transformation $\Pzero$,
\begin{equation}
\begin{split}
\label{eqn:BdG_s-wave-z-1a}
\A_0 &= {\Pzero}^{-1} (-\mathcal{M}_0) \Pzero = \begin{pmatrix}
 0 & 1 & 0 & 0 \\
 0 & 0 & 0 & 0 \\
 0 & 0 & 0 & 0 \\
 0 & 0 & 0 & 0 \\
\end{pmatrix}
, \\
\Pzero &= \begin{pmatrix}
 0 & 0 & 0 & 1 \\
 0 & 0 & 1 & 0 \\
 0 & -1 & 0 & 0 \\
 1 & 0 & 0 & 0 \\
\end{pmatrix},
\end{split}
\end{equation}
where for definiteness we have fixed $\ell = +1$. $\A_0$ is now in the canonical Jordan block diagonal form
\begin{equation}
\label{eqn:BdG_s-wave-z-4}
\A_0 = \begin{pmatrix}
H_1& 0 & 0 \\
0 &H_2 & 0 \\
0 & 0 & H_3 \\
\end{pmatrix} = H_1  \oplus H_2 \oplus H_3  ,
\end{equation}
where $H_j$ ($j = 1,2,3$) are shifting matrices where $H_1$ is of dimension two and $H_2$, $H_3$ are of dimension 1. The same transformation is applied to all the matrices, defining the new starting point $\A = {\Pzero}^{-1}(- \mathcal{M}) \Pzero$, $ X \equiv \Pzero Y$ such that
\begin{equation}
\label{eqn:BdG_s-wave-z-1b}
x^{-q}Y^\prime(x) = \A(x) Y(x),
\end{equation}
where $q =0$, and 
\begin{subequations}
\label{eqn:BdG_s-wave-z-1c-sys}
 \begin{align}
\label{eqn:BdG_s-wave-z-1c-a}
 \A_0 &= \begin{pmatrix}
 0 & 1 & 0 & 0 \\
 0 & 0 & 0 & 0 \\
 0 & 0 & 0 & 0 \\
 0 & 0 & 0 & 0 \\
\end{pmatrix} ,\\
\label{eqn:BdG_s-wave-z-1c-b}
\A_1 &=  \begin{pmatrix}
 1 & \frac{f^2}{\lambda } & 0 & 0\\
 0 & 0 & 0 & 0 \\
 0 & \frac{2m \lambda }{\hbar ^2} & 1 & -\frac{f^2}{\lambda } \\
 0 & 0 & 0 & 0 \\
 \end{pmatrix} ,\\
\label{eqn:BdG_s-wave-z-1c-c}
\A_2 &=  \begin{pmatrix}
 -\frac{2 f^2}{\lambda } & \frac{2 m (h-\mu )}{\hbar ^2}-\frac{f^4}{\lambda ^2} & \frac{2 m \lambda }{\hbar ^2} & 0 \\
 1 & 0 & 0 & 0 \\
 -\frac{2 m \lambda }{\hbar ^2} & 0 & -\frac{2 f^2}{\lambda } & \frac{f^4}{\lambda ^2}+\frac{2 m (h+\mu )}{\hbar ^2} \\
 0 & 0 & -1 & 0 \\
\end{pmatrix}.
 \end{align}
\end{subequations}

\subsubsection{\label{Sec:PrepR1}Prepare for a shearing transformation that simplifies the problem}
The transformation $Y(x) = \K(x)Z(x)$, where the matrix $\K(x)$ is holomorphic and has a non-vanishing determinant at $x = \infty$ changes Eq.~\eqref{eqn:BdG_s-wave-z-1b} into
\begin{equation}
\label{eqn:BdG_s-wave-z-2}
x^{-q}Z^\prime(x) = \B(x)  Z(x)
\end{equation}
with $q \geq 0$ and
\begin{equation}
\label{eqn:BdG_s-wave-z-3}
x^{-q}\K^\prime(x) = \A(x) \K(x)   - \K(x) \B(x).
\end{equation}
We seek series solutions $\K(x) =  \sum_{\nu = 0}^\infty \K_\nu x^{-\nu}$, $\B(x) =  \sum_{\nu = 0}^\infty \B_\nu x^{-\nu}$, where we recall $\A(x) = \sum_{\nu = 0}^\infty \A_\nu x^{-\nu}$.  We set
\begin{subequations}
\label{eqn:BdG_s-wave-z-5-sys}
 \begin{align}
 \B_0 = \A_0, \\
 \K_0 = I.
 \end{align}
\end{subequations}
Using Eq.~\eqref{eqn:BdG_s-wave-z-5-sys}, substitution of the series ans\"atze into Eq.~\eqref{eqn:BdG_s-wave-z-3} and comparison of like coefficients results in the recursion relation
\begin{subequations}
\label{eqn:BdG_s-wave-z-6-sys}
 \begin{align}
 \label{eqn:BdG_s-wave-z-6-a}
\A_0 \K_0 - \K_0 \A_0 &= 0, \\
\label{eqn:BdG_s-wave-z-6-b}
\A_0 \K_\nu - \K_\nu \A_0 &=\sum_{s = 0}^{\nu - 1} (\K_s \B_{\nu - s} - \A_{\nu  - s} \K_s) \\
& \notag - (\nu - q - 1)\K_{\nu - q- 1} \qquad (\nu > 0),
 \end{align}
\end{subequations}
where the last term in Eq.~\eqref{eqn:BdG_s-wave-z-6-b} is absent for $\nu - q- 1 < 0$, that is, for $\nu < 1$. The recursion relation is of the form
\begin{equation}
 \label{eqn:BdG_s-wave-z-7}
\A_0 \K_\nu - \K_\nu \A_0 = \B_\nu  + K_\nu, \qquad \nu > 0
\end{equation}
where $K_\nu =\sum_{s = 1}^{\nu - 1} \K_s \B_{\nu - s} - \sum_{s = 0}^{\nu - 1}  \A_{\nu  - s} \K_s - (\nu - q - 1)\K_{\nu - q- 1}$ depends only on the $\K_j$, $\B_j$ with $j < \nu$. 
If all the eigenvalues of $\A_0$ are distinct, then $\B(x)$ will be diagonal and the problem is uncoupled and easily solved. If at least two eigenvalues are distinct, we can take all $\B_\nu$ ($\nu > 0$) zero or block-diagonal. However, this is not possible here because $\A_0$ has only one distinct eigenvalue, zero.

Instead, we partition each Eq.~\eqref{eqn:BdG_s-wave-z-7} into blocks of the same order as the Jordan blocks $H_j$ for $\A_0$, and call these blocks $\K_\nu^{jk}$  with $j,k = 1,2,\ldots, s$. For us $s = 3$. Then each relation~\eqref{eqn:BdG_s-wave-z-7}  corresponds to $s^2 = 9$ relations  
\begin{equation}
 \label{eqn:BdG_s-wave-z-8}
H_j \K_\nu^{jk} - \K_\nu^{jk} H_k = \B_\nu^{jk}  + K_\nu^{jk}, \qquad \nu > 0
\end{equation}
It can be proven that the equation $AX - XB = 0$ possesses solutions other than $X = 0$ if and only if $A$ and $B$ have at least one common eigenvalue. Since this result guarantees the existence of non-trivial solutions to the corresponding homogeneous equations of Eq.~\eqref{eqn:BdG_s-wave-z-8},  each equation of Eq.~\eqref{eqn:BdG_s-wave-z-8} can be soluble only if the matrices $\B_\nu^{jk}$ satisfy some restrictive condition, explained below. Consider the auxiliary equation
\begin{equation}
 \label{eqn:BdG_s-wave-z-9}
HX - XK = M,
\end{equation}
where $H$ and $K$ are shifting matrices of orders $h$ and $k$ respectively, and $M$ is a $h \times k$ matrix whose first $h - 1$ rows are given constant vectors while the entries $\alpha_1, \alpha_2, \ldots, \alpha_k$ of the last row are variables. It can be proven that the numbers  $\alpha_1, \alpha_2, \ldots, \alpha_k$ can be determined uniquely in such a way that Eq.~\eqref{eqn:BdG_s-wave-z-9} is solvable for the $h \times k$ matrix $X$. We now apply this result to Eq.~\eqref{eqn:BdG_s-wave-z-8}. Without loss of generality we take all rows but the last of $\B_\nu^{jk}$ to be zero. Then, the series
\begin{equation}
\sum_{\nu = 0}^\infty \K_\nu x^{-\nu}
\end{equation}
is determined by solving the recursion relations~\eqref{eqn:BdG_s-wave-z-8} successively for $\nu = 1, 2, \ldots$. While it will in general be divergent, it can be proven that it is the asymptotic expansion of some holomorphic matrix function $\K(x)$ for sufficiently large $x_0$ such that $|x| > x_0$.

The point of the transformation $\K$ is to obtain a differential equation~\eqref{eqn:BdG_s-wave-z-2} such that the matrices $\B$, by construction,  satisfy the following three properties: (i) $\B(x) = \sum_{\nu = 0}^\infty \B_\nu x^{-\nu}$; (ii) $\B_0 = H_1  \oplus H_2 \oplus \cdots \oplus H_s $ (the $H_j$ are shifting matrices); and, most importantly, (iii) that the only non-zero entries in $\B_\nu$ for $\nu > 0$ occur in the rows corresponding to the last rows of the Jordan blocks $H_k$ ($k = 1, 2, \ldots, s$) in the representation of $\A_0$. The transformation $\K$ induces zeros into the matrices preparing them for a shearing transformation. The result is an asymptotic series solution valid at $x \to \infty$ such that $\B(x) = \B_0 + \B_1 x^{-1}  + \B_2 x^{-2} + \B_3 x^{-3} + \ldots$, where
\begin{subequations}
 \label{eqn:BdG_s-wave-z-9a-sys}
 \begin{align}
 \label{eqn:BdG_s-wave-z-9a-sys-a}
  \B_0 &=  \begin{pmatrix}
 0 & 1 & 0 & 0 \\
 0 & 0 & 0 & 0 \\
 0 & 0 & 0 & 0 \\
 0 & 0 & 0 & 0 \\
\end{pmatrix} ,
\qquad
  \B_1 =  \begin{pmatrix}
 0 & 0 & 0 & 0 \\
 0 & 1 & 0 & 0 \\
 0 & \frac{2 m \lambda }{\hbar ^2} & 1 & -\frac{f^2}{\lambda } \\
 0 & 0 & 0 & 0 \\
\end{pmatrix} , \\
 \label{eqn:BdG_s-wave-z-9a-sys-b}
\B_2 &=  \begin{pmatrix}
 0 & 0 & 0 & 0 \\
 0 & -\frac{3 f^2}{\lambda } & 0 & 0 \\
 -\frac{4 m \lambda }{\hbar ^2} & -\frac{2 f^2 m}{\hbar ^2} & -\frac{2 f^2}{\lambda } & \frac{f^4}{\lambda ^2}+\frac{2 m (h+\mu )}{\hbar ^2} \\
 0 & 0 & -1 & 0 \\
 \end{pmatrix}, \\
 \label{eqn:BdG_s-wave-z-9a-sys-c}
 \B_3 &=  \begin{pmatrix}
 0 & 0 & 0 & 0 \\
 \frac{6 f^2}{\lambda } & \frac{3 f^4}{\lambda ^2}+\frac{4 m \left(m \lambda ^2+(\mu -h) \hbar ^2\right)}{\hbar ^4} & -\frac{2 m \lambda }{\hbar ^2} & -\frac{2 f^2 m}{\hbar ^2} \\
 \frac{6 f^2 m}{\hbar ^2} & \frac{4 m \left(\hbar ^2 f^4-h m \lambda ^2+m \lambda ^2 \mu \right)}{\lambda  \hbar ^4} & -\frac{4 m^2 \lambda ^2}{\hbar ^4} & 0 \\
 0 & 0 & 0 & 0 \\
 \end{pmatrix}, \\
 \B_4 &=   \cdots .
 \end{align}
\end{subequations}
We work until $\nu = \nmax$. The final series solutions will then be exact upto $\nu = \nmax - 1$ for $F_\ua(r)$ and $\nu = \nmax - 2$ for $F_\da(r)$. For spin-$\ua$ the derivative (component 2) will agree with a direct derivative of component 1 upto $\nu = \nmax - 2$ and for spin-$\da$ the derivative (component 4) will agree with a direct derivative of component 3 upto $\nu = \nmax - 3$. 

\subsubsection{Simplify the problem by a shearing transformation}
Having obtained the matrix $\B$, Eq.~\eqref{eqn:BdG_s-wave-z-2} is ready for the shearing transformation
\begin{equation}
\label{eqn:BdG_s-wave-z-10}
Z = S(x) V \equiv \mathrm{diag}(1,x^{-\xi}, x^{-2\xi},x^{-3\xi})  V
\end{equation}
with a temporarily unknown positive parameter $\xi$, which takes Eq.~\eqref{eqn:BdG_s-wave-z-2} into
\begin{equation}
\label{eqn:BdG_s-wave-z-11}
x^{-q} V^\prime(x) = \C V =  \left(\sum_{\nu = 0}^\infty \C_\nu x^{-\nu} \right) V,
\end{equation}
where the matrix $\C(x) = S^{-1}(x) \B(x) S(x) - x^{-q}S^{-1}(x)S^\prime (x)$. The matrix elements $c_{jk}$ ($j,k = 1, \ldots, 4$, $n = 4$) of $\C$ read
\begin{equation}
\label{eqn:BdG_s-wave-z-12}
c_{jk} = b_{jk} x^{\xi(j-k)} + (j - 1)\xi \delta_{jk} x^{-q-1}.
\end{equation}
Here $\delta_{jk}$ is the Kronecker delta, and $b_{jk}$ are the matrix elements of $\B$. 
The parameter $\xi$ must be chosen appropriately to induce, where possible, non-zero elements below the main diagonal into the leading order matrix $\C_0$.  Above the main diagonal it is equal to $\B_0$. 

The parameter $\xi$ must be chosen judiciously as follows. Any $b_{jk} \neq 0$ is of the form
\begin{equation}
\label{eqn:BdG_s-wave-z-13}
b_{jk} = x^{-\alpha_{jk}} \sum_{\nu =0 }^\infty b_{jk \nu} x^{-\nu},
\end{equation}
where $b_{jk \nu} \neq 0$ and each positive integer $\alpha_{jk} \geq 1$ except for the special elements with a 1 from a shifting matrix in the Jordan decomposition for $\B_0$ for which $\alpha_{jk} = 0$. Since we could not diagonalise $\B_0$ and instead obtained a Jordan matrix, at least one such special element is present. Before the shearing transformation ($\xi = 0$) the special elements have the lowest $\alpha_{jk}$, and the purpose of the transformation is to add non-zero elements on or below the diagonal to $\B_0$ by a suitable choice of $\xi$. After the shearing transformation the expansions of non-zero off-diagonal elements $c_{jk}$ ($\delta_{jk} = 0$) begin with the power $-\alpha_{jk} + \xi(j-k)$. The special elements on the superdiagonal begin with the power $x^{-\xi}$. There exists a smallest rational $\xi_0 = q\a/p\a > 0$ with $q\a, p\a$ coprime, for which the special elements have the same leading power as an element below the main diagonal, $\xi_0 = \alpha_{jk} - \xi_0(j-k)$ for some $j,k<j$. We take $\xi = \xi_0$ in the shearing transformation; the result of the above process is $\xi = q\a/p\a$ with $q\a = 1$, $p\a = 2$ coprime. 
Fractional powers thus unavoidably appear in the shearing transformation~\eqref{eqn:BdG_s-wave-z-10}.

In descending powers of $x^{-1/p\a}$, the matrix $\C$ will begin with the power $x^{-q\a/p\a}$. The matrix $x^\xi \C$ as $x \to \infty$ has at least one non-zero entry on or below the main diagonal and above the main diagonal it is equal to $\B_0$:
\begin{equation}
 \label{eqn:BdG_s-wave-z-13a}
 \lim_{x\to \infty} x^\xi \C(x) =  \begin{pmatrix}
 0 & 1 & 0 & 0 \\
 0 & 0 & 0 & 0 \\
 0 & \frac{2 m \lambda }{\hbar ^2} & 0 & 0 \\
 0 & 0 & 0 & 0 \\
\end{pmatrix} .
 \end{equation}
We remove the fractional powers by multiplying both sides by $x^\xi$ and introducing the new independent variable $x_1$ such that
\begin{equation}
\label{eqn:BdG_s-wave-z-14}
x =  {p\a}^{1/(\xi - q - 1)} x_1^{p\a} = x_1^2/4,
\end{equation}
to obtain from Eq.~\eqref{eqn:BdG_s-wave-z-11} the differential equation
\begin{equation}
\label{eqn:BdG_s-wave-z-15}
x_1^{-h} V^\prime(x_1) = \D(x_1) V =  \left(\sum_{\nu = 0}^\infty \D_\nu x_1^{-\nu} \right) V,
\end{equation}
where
\begin{subequations}
\label{eqn:BdG_s-wave-z-16-sys}
 \begin{align}
 \label{eqn:BdG_s-wave-z-16-a}
h &= p\a q + p\a -p\a \xi  - 1 = 0,\\
 \label{eqn:BdG_s-wave-z-16-b}
d_{ij} &= \left({p\a}^{1/(\xi - q - 1)} t^{p\a}  \right)^\xi \left[ b_{jk} {p\a}^{\xi(j-k)/(\xi - q - 1)} x_1^{q\a(j-k)} \right. \\
&\notag \left. \qquad + (j - 1) \frac{q\a}{p\a} \delta_{jk} {p\a}^{(-q-1)/(\xi - q - 1)} x_1^{p\a(-q-1)}\right] \\
\notag &= b_{jk} \left(\frac{x_1}{2}\right)^{j-k+1}  + \delta_{jk}(j - 1)  x_1^{-1}.
 \end{align}
\end{subequations}
The branch of the multi-valued function $x^{p\a}$ can be chosen freely.

While new elements have appeared on and below the main diagonal in $\D_0$, the upper triangular part above the main diagonal, in particular the superdiagonal, of $\D_0$ is equal to that of $\B_0$ by construction. This completes the purpose of the shearing transformation.

If $h < 0$, the problem would be solved because then either the singular point would be regular ($h = -1$), or there would not be any singular point. If $h \geq 0 $ and $\D_0$ has at least two distinct eigenvalues, then the problem reduces to a set of similar problems of lower order. However, the eigenvalues of $\D_0$ are all equal to zero, and $\D_0$ is nilpotent. It can be proven~\cite{wasow2002asymptotic} that if the process outlined above is repeatedly carried out, eventually we obtain a nilpotent $\D_0$ that is itself a shifting matrix i.e. $s = 1$.

\subsubsection{Obtain an equation with a regular singular point}
The entire process must be reapplied, starting from the JCF for $\D_0$. Another shearing transformation with the parameter $1/2$ is required. Repeating the process once more, the third shearing transformation comes with an integer-valued shearing parameter of $1$, which means that via such a chain of transformations we have reduced the coupled system~\eqref{eqn:BdG_s-wave-z-1b},
\begin{equation}
Y^\prime(x) = \A(x) Y(x) = \left(\sum_{\nu = 0}^\infty \A_\nu x^{-\nu} \right) Y,
\end{equation}
to the system
\begin{equation}
 \label{eqn:BdG_s-wave-z-45}
x_2 {Y\c}^\prime(x_2) = \Ac Y\c =  \left(\sum_{\nu = 0}^\infty \Ac_\nu x^{-\nu} \right) Y\c,
\end{equation}
where 
\begin{equation}
 \label{eqn:BdG_s-wave-z-46}
 \Ac_0 = \begin{pmatrix}
 1 & 0 & 0 & 0 \\
 0 & 9 & 1 & 0 \\
 0 & 0 & 9 & 0 \\
 0 & 0 & 0 & 13 \\
\end{pmatrix},
\qquad
\Ac_1 =  \cdots ,
\end{equation}
and $x_1 = x_2^2/4$. 
We have reached our goal that we set below Eq.~\eqref{eqn:BdG_s-wave-z-1}: Eq.~\eqref{eqn:BdG_s-wave-z-45} has only a regular singular point at infinity (or the vortex core in terms of $r$). This equation can be solved with more standard methods.

Through a sequence of standard double transformations that raise the eigenvalues of a Jordan block of $\Ac_0$ by an integer followed by a non-singular constant similarity transformation of $\Ac_0$ to JCF, we obtain 
\begin{equation}
 \label{eqn:BdG_s-wave-z-55-a-5}
x_2 {Y\f}^\prime(x_2) = \Af Y\f 
\end{equation}
such that no eigenvalues of the leading matrix $\Af_0$ differ by a positive integer. In fact, it consists of two identical Jordan blocks: $\Af_0 = \begin{pmatrix}
 13 & 1  \\
 0 & 13 \\
\end{pmatrix} \oplus  \begin{pmatrix}
 13 & 1  \\
 0 & 13 \\
\end{pmatrix}$.

Let us now transform back to radial coordinates with $r_2 = 1/x_2$ (here $x_1 = x_2^2/4$) to obtain from Eq.~\eqref{eqn:BdG_s-wave-z-55-a-5} the equation
\begin{equation}
 \label{eqn:BdG_s-wave-z-56}
r_2 {Y\g}^\prime(r_2) = \Ag(r_2) Y\g=   \left( \sum_{\nu = 0}^\infty \Ag_\nu r_2^{\nu}\right) Y\g,
\end{equation}
where $\Ag(r_2) = -\Af(1/r_2)$ and $\Ag(0)$ is holomorphic.

Since $\Ag(r_2)$ is holomorphic at $r_2 =0$ and since no two eigenvalues of $\Ag_0$ differ by a positive integer, Eq.~\eqref{eqn:BdG_s-wave-z-56} has a fundamental matrix solution of the form
\begin{equation}
 \label{eqn:BdG_s-wave-z-57}
Y\g(r_2) = \Kg(r_2) \Upsilon(r_2), \qquad  \Kg_0 = I,
\end{equation}
where $\Kg(0)$ is holomorphic. Its power series representation $\Kg =   \sum_{\nu = 0}^\infty \Kg_\nu r_2^{\nu}$ can be calculated by rational operations from the coefficients $\Ag_\nu$ in the series $\Ag= \sum_{\nu = 0}^\infty \Ag_\nu r_2^{\nu}$. 

The matrix $\Kg(r_2)$ in the transformation $Y\g = \Kg(r_2)  \Upsilon$, which results in the equation
\begin{equation}
 \label{eqn:BdG_s-wave-z-58}
r_2  \Upsilon^\prime(r_2) = \Bg(r_2)  \Upsilon =   \left( \sum_{\nu = 0}^\infty \Bg_\nu r_2^{\nu}\right)  \Upsilon,
\end{equation}
 is calculated using the recursion relation~\eqref{eqn:BdG_s-wave-z-6-sys} with $q = -1$ and $r = 1/x$ (also changing the sign of the last term in Eq.~\eqref{eqn:BdG_s-wave-z-59-b}), viz.
\begin{subequations}
\label{eqn:BdG_s-wave-z-59-sys}
 \begin{align}
 \label{eqn:BdG_s-wave-z-59-a}
\Ag_0 \Kg_0 - \Kg_0 \Ag_0 &= 0, \\
\label{eqn:BdG_s-wave-z-59-b}
\Ag_0 \Kg_\nu - \Kg_\nu \Ag_0 &=\sum_{s = 0}^{\nu - 1} (\Kg_s \Bg_{\nu - s} - \Ag_{\nu  - s} \Kg_s) \\
& \notag \qquad + \nu \Kg_{\nu} \qquad (\nu > 0),
 \end{align}
\end{subequations}
where
\begin{subequations}
\label{eqn:BdG_s-wave-z-60-sys}
 \begin{align}
 \Bg_0 &= \Ag_0, \\
 \Kg_0 &= I.
 \end{align}
\end{subequations}
Given that the matrix $\Ag_0$ in the convergent expansion $\Ag= \sum_{\nu = 0}^\infty \Ag_\nu r_2^{\nu}$ has no eigenvalues that differ from each other by positive integers, there exists a formal convergent series $\Kg(r_2)$ with $\Kg_0 = I$ such that the formal transformation $Y\g = \Kg(r_2) \Upsilon$  reduces the differential equation~\eqref{eqn:BdG_s-wave-z-56} to the form of Eq.~\eqref{eqn:BdG_s-wave-z-58} such that $\Bg_\nu = 0$ for all $\nu > 0$. Therefore, the recursion relations~\eqref{eqn:BdG_s-wave-z-59-sys} can be solved for $\Kg(r_2)$ such that $\Bg_\nu = 0$ for $\nu > 0$. The solution for $Y\g(r_2)$ is then given by Eq.~\eqref{eqn:BdG_s-wave-z-57}.

Since $\left[\frac{\Bg_0}{r_1}, \frac{\Bg_0}{r_2}\right] = 0$, the general solution to the equation 
\begin{equation}
 \label{eqn:BdG_s-wave-z-61}
r_2  \Upsilon^\prime(r_2) = \Bg_0  \Upsilon
\end{equation}
is given by the matrix exponential~\eqref{eqn:BdG_s-wave-t1-2},
\begin{equation}
 \label{eqn:BdG_s-wave-z-62}
 \Upsilon(r_2)  =  \Upsilon(r_0)\, \mathrm{exp}\left\lbrace \Bg_0 \ln{\left( \frac{r_2}{r_0}\right)} \right\rbrace.
\end{equation}
Here a set of four linearly independent vectors forms the fundamental matrix solution $ \Upsilon(r_2)$, and $ \Upsilon(r_0)$ is the fundamental matrix solution at $r_2 = r_0$. We can set the auxiliary parameter $r_0 = 1$ and choose $ \Upsilon(r_0)$ in such a way as to choose the boundary conditions for $F_\sigma(r)$ at $r = 0$. The general (vector) solution is then given by linear combinations of the columns.

The logarithms of the fundamental solution matrix end up in columns 2 and 4 making them either divergent or zero at the origin $r = 0$. The columns 1 and 3, in contrast, can be chosen to remain well-behaved and finite at the vortex core $r = 0$. We do not include the columns 2 and 4 in what follows, that is, we take a linear combination only of columns 1 and 3.

Performing all the transformations carried out in an inverse order, we can compute the matrix solution $X$ for the original problem~\eqref{eqn:BdG_s-wave-4-simplf-aux-sys} from knowing $\Upsilon(r_2)$. A judicious choice for $\Upsilon(1)$ (with $r_0 = 1$) then gives the explicit solution shown in Eq.~\eqref{eqn:BdG_s-wave-SOLUTION} in the main text.

\end{document}